\documentclass{elsart}
\usepackage{epsfig,cite}
\usepackage{color}

\begin{document}
\newcommand{\aff}[2]{Dipartimento di Fisica dell'Universit\`a #1 e Sezione
  INFN, #2, Italy.} 
\newcommand{\affd}[1]{Dipartimento di Fisica dell'Universit\`a e Sezione
  INFN, #1, Italy.}
\def\etc{{\it etc.}} \def\eg{{\it e. g.}}  \let\cl=\centerline
\def\ie{{\it\kern-2pt i.\kern-.5pt e.\kern-2pt}}  \def\etal{{\it et al.}}
\def\BR{\hbox{BR}}  \def\prl{Phys. Rev. Lett}
\def\up#1{$^{#1}$}  \def\dn#1{$_{#1}$}
\def\ifm#1{\relax\ifmmode#1\else$#1$\fi}
\def\to{\ifm{\rightarrow}} \def\sig{\ifm{\sigma}}   \def\plm{\ifm{\pm}}
\def\K{\ifm{K}} \def\LK{\ifm{L_K}} 
\def\ff{$\phi$--factory}  \def\DAF{DA\char8NE}  \def\f{\ifm{\phi}} 
\def\pb{{\bf p}} \def\pic{\ifm{\pi^+\pi^-}} \def\pio{\ifm{\pi^0\pi^0}} 
\def\gam{\ifm{\gamma}} \def\kkb{\ifm{\ko\kob}} 
\def\dt{ \ifm{{\rm d}t} } \def\ab{\ifm{\sim}}  \def\x{\ifm{\times}}
\def\sta#1,{\ifm{|\,#1\,\rangle}} \def\ket#1,{\ifm{|\,#1\,\rangle}} 
\def\L{\ifm{{\cal L}}}  \def\R{\ifm{{\cal R}}}
\def\pt#1,#2,{\ifm{#1\x10^{#2}}}
\def\kon{\ifm{K_1}} \def\ktw{\ifm{K_2}} 
\def\minus{$-$}  \def\dif{\hbox{d}}   \def\Gam{\ifm{\Gamma}}
\def\ktagmuii{\ifm{K^{-}\to\mu^{-}\nu}}
\def\kmu{\ifm{K\to\mu\nu(\gamma)}}
\def\kmuii{\ifm{K^{+}\to\mu^{+}\nu(\gamma)}}
\def\kmuiig{\ifm{K^{+}\to\mu^{+}\nu\gamma}}
\def\pmuii{\ifm{\pi^+\to\mu^+\nu(\gamma)}}
\def\vus{\ifm{V_{us}}}   \def\vud{\ifm{V_{ud}}}
\def\kpm{\ifm{K^\pm}}  \def\epm{\ifm{e^+e^-}}
\def\kppipi{\ifm{K^{+}\to \pi^{+}\pi^0}}
\def\kppil{\ifm{K^{+} \to \pi^0 l^{+}\nu}}
\def\kpmunu{\ifm{K^{+} \to \mu^{+}\nu}}
\def\figbox#1;#2;{\parbox{#2cm}{%
\vglue3mm\epsfig{file=\figdir#1.eps,width=#2cm}\vglue3mm}}
\def\figboxc#1;#2;{\cl{\figbox #1;#2;}}
\def\figdir{}
\def\kp{\ifm{K^+}} \def\km{\ifm{K^-}}
\def\bye{\end{document}}
\begin{frontmatter}
\title{Measurement of the Absolute Branching Ratio for the $K^+\rightarrow\mu^+\nu(\gamma)$ Decay with the KLOE Detector.}
\collab{The KLOE Collaboration}
\author[Na]{F.~Ambrosino},
\author[Frascati]{A.~Antonelli},
\author[Frascati]{M.~Antonelli},
\author[Roma3]{C.~Bacci},
\author[Frascati]{P.~Beltrame},
\author[Frascati]{G.~Bencivenni},
\author[Frascati]{S.~Bertolucci},
\author[Roma1]{C.~Bini},
\author[Frascati]{C.~Bloise},
\author[Roma1]{V.~Bocci},
\author[Frascati]{F.~Bossi},
\author[Frascati,Virginia]{D.~Bowring},
\author[Roma3]{P.~Branchini},
\author[Roma1]{R.~Caloi},
\author[Frascati]{P.~Campana},
\author[Frascati]{G.~Capon},
\author[Na]{T.~Capussela},
\author[Roma3]{F.~Ceradini},
\author[Frascati]{S.~Chi},
\author[Na]{G.~Chiefari},
\author[Frascati]{P.~Ciambrone},
\author[Virginia]{S.~Conetti},
\author[Frascati]{E.~De~Lucia\thanksref{*}},
\author[Frascati]{P.~De~Simone},
\author[Roma1]{G.~De~Zorzi},
\author[Frascati]{S.~Dell'Agnello},
\author[Karlsruhe]{A.~Denig},
\author[Roma1]{A.~Di~Domenico},
\author[Na]{C.~Di~Donato},
\author[Pisa]{S.~Di~Falco},
\author[Roma3]{B.~Di~Micco},
\author[Na]{A.~Doria},
\author[Frascati]{M.~Dreucci},
\author[Frascati]{G.~Felici},
\author[Karlsruhe]{A.~Ferrari},
\author[Frascati]{M.~L.~Ferrer},
\author[Frascati]{G.~Finocchiaro},
\author[Frascati]{C.~Forti},
\author[Roma1]{P.~Franzini},
\author[Frascati]{C.~Gatti},
\author[Roma1]{P.~Gauzzi},
\author[Frascati]{S.~Giovannella},
\author[Lecce]{E.~Gorini},
\author[Roma3]{E.~Graziani},
\author[Pisa]{M.~Incagli},
\author[Karlsruhe]{W.~Kluge},
\author[Moscow]{V.~Kulikov},
\author[Roma1]{F.~Lacava},
\author[Frascati]{G.~Lanfranchi},
\author[Frascati,StonyBrook]{J.~Lee-Franzini},
\author[Karlsruhe]{D.~Leone},
\author[Frascati]{M.~Martini},
\author[Na]{P.~Massarotti},
\author[Frascati]{W.~Mei},
\author[Na]{S.~Meola},
\author[Frascati]{S.~Miscetti},
\author[Frascati]{M.~Moulson},
\author[Karlsruhe]{S.~M\"uller},
\author[Frascati]{F.~Murtas},
\author[Na]{M.~Napolitano},
\author[Roma3]{F.~Nguyen},
\author[Frascati]{M.~Palutan},
\author[Roma1]{E.~Pasqualucci},
\author[Roma3]{A.~Passeri},
\author[Frascati,Energ]{V.~Patera},
\author[Na]{F.~Perfetto},
\author[Roma1]{L.~Pontecorvo},
\author[Lecce]{M.~Primavera},
\author[Frascati]{P.~Santangelo},
\author[Roma2]{E.~Santovetti},
\author[Na]{G.~Saracino},
\author[Frascati]{B.~Sciascia},
\author[Frascati,Energ]{A.~Sciubba},
\author[Pisa]{F.~Scuri},
\author[Frascati]{I.~Sfiligoi},
\author[Frascati]{T.~Spadaro},
\author[Roma1]{M.~Testa},
\author[Roma3]{L.~Tortora},
\author[Roma1]{P.~Valente},
\author[Karlsruhe]{B.~Valeriani},
\author[Frascati]{G.~Venanzoni},
\author[Roma1]{S.~Veneziano},
\author[Lecce]{A.~Ventura},
\author[Karlsruhe]{R.Versaci\thanksref{*}},
\author[Frascati,Beijing]{G.~Xu}
%
\address[Beijing]{Permanent address: Institute of High Energy Physics of
  Academica Sinica, Beijing, China.}
\address[Frascati]{Laboratori Nazionali di Frascati dell'INFN, Frascati, Italy.}
\address[Karlsruhe]{Institut f\"ur Experimentelle Kernphysik, Universit\"at
  Karlsruhe, Germany.}
\address[Lecce]{\affd{Lecce}}
\address[Moscow]{Permanent address: Institute for Theoretical and
  Experimental Physics, Moscow, Russia.}
\address[Na]{Dipartimento di Scienze Fisiche dell'Universit\`a ``Federico
  II'' e Sezione INFN, Napoli, Italy}
\address[Novo]{Permanent address: Budker Institute of Nuclear Physics,
  Novosibirsk, Russia.}
\address[Pisa]{\affd{Pisa}}
\address[Energ]{Dipartimento di Energetica dell'Universit\`a ``La
  Sapienza'', Roma, Italy.}
\address[Roma1]{\aff{``La Sapienza''}{Roma}}
\address[Roma2]{\aff{``Tor Vergata''}{Roma}}
\address[Roma3]{\aff{``Roma Tre''}{Roma}}
\address[StonyBrook]{Physics Department, State University of New York at Stony Brook, USA.}
\address[Tbilisi]{Permanent address: High Energy Physics Institute, Tbilisi
  State University, Tbilisi, Georgia.}
\address[Virginia]{Physics Department, University of Virginia, USA.}
\thanks[*]{Corresponding authors.\\
           {\it e-mail addresses:} erika.delucia@lnf.infn.it (E. De
           Lucia),\\ versaci@iekp.fzk.de (R. Versaci).}
%
\begin{abstract}
We have measured the fully inclusive \kmuii\ absolute branching ratio with
the KLOE experiment at \DAF, the Frascati \ff. From some 865,283 \kmuii\
decays obtained from a sample of \ab\pt5.2,8, \f-meson decays, we find
BR($K^+\to\mu^+\nu_\mu (\gam))$=0.6366\plm0.0009\dn{\rm
  stat.}\plm0.0015\dn{\rm syst.}, corresponding to an overall fractional
error of 0.27\%. Using recent lattice results on the decay constants of
pseudoscalar mesons one can obtain an estimate for the CKM mixing matrix
element $|V_{us}|$=0.2223\plm0.0026. 
\end{abstract}
%
%
\end{frontmatter}
%
%
\section{Introduction} 
The most recent measurement of the  $K\to\mu\nu$  branching ratio,
ref. \citen{chiang}, based on 62,000 events, dates back to 1972, more than
30 years ago, and relies on a sample of \ab10\up5 kaon decays. The authors
of ref. \citen{chiang} 
quote an error of \ab0.7\%, the statistical error due to the event count
being 0.4\%. This error is reduced in the PDG fit \cite{pdg} to 0.27\% and
the value changed by 0.3\%. While the procedure is correct in principle, it can lead
to incorrect results because of incorrect data included in the fit. This
has been the case sometimes, leading among other things to the suggestion,
also in ref. \citen{pdg}, of a value for $|V_{us}|$ apparently inconsistent
with unitarity of the CKM mixing matrix \cite{somref}.
Another problem with measurements performed more than 30 years ago is due
to the fact that the effect 
of radiative corrections was not fully
appreciated. It is therefore impossible to understand what fraction of the
radiative decay is included in the quoted results. Inclusion of all
radiation is however necessary to compare to models or to extract
fundamental parameters such as a coupling constant. 
With all the above in mind as well as recent developments in numerical or
lattice QCD calculation, we have begun a program of new precise, fully
inclusive, kaon branching ratio measurement. \\[5mm] 
We report in the following a measurement of BR(\kmuii) performed with the
KLOE detector at \DAF, the Frascati \ff. The measurement is based on an
integrated luminosity of \ab175 pb\up{-1}, collected in 2001-02. \DAF\ is
an \epm\ collider  
operated at a total energy of $W=1020$ MeV, the mass of the
\f(1020)-meson. Equal energy positron and electron beams collide at an
angle of ($\pi-$)25 mrad and  
produce \f-mesons with a small transverse momentum of \ab12.5 MeV/c. The
collision frame moves therefore in the laboratory with a velocity
$\beta$\ab0.0125. 
In the center of mass, the \f-meson decays into anti-collinear
\kpm\ pairs of \ab125 MeV/c momentum. In the laboratory this remains
approximately true: detection of a $K^\pm$ tags the presence of a $K^\mp$
of given momentum and direction.
The decay products of the \kpm\ pair define two spatially well separated
regions called in the following the tag and the signal hemispheres. 
Identified $K^\mp$ decays tag a $K^\pm$ beam and provide an absolute count.
This procedure is a unique feature of a $\phi$-factory and provides the
means for measurements of absolute branching ratios, i.e. ratios
$\Gamma_{i}/\Gamma_{tot}$ rather than ratios of BR's $\Gamma_{i}/\Gamma_{j}$.
%
\section{The KLOE detector}
The KLOE detector consists of a large volume drift chamber and a sampling
calorimeter. The drift chamber (DC) \cite{DC}, of 3.3 m length and 2 m
radius, has a full stereo geometry and operates with a 90\% helium-10\%
isobutane gas mixture. Tracking in the DC provides measurements of the
vector momentum of charged particles with
$\sigma(p_{\perp})/p_{\perp}\leq0.4\%$ and two track
vertices to 3 mm. In the following we use a coordinate system with the
$z$-axis defined as the bisectrix of the \epm\ beams, the $y$-axis vertical
and the $x$-axis toward the center of the collider rings.
\\[5mm]
The calorimeter (EMC) \cite{EMC} consists of a cylindrical 
barrel and two endcaps
covering a solid angle of 98\% of 4$\pi$. Photons showering in the
lead-scintillator-fiber EMC structure are detected as local energy deposits
by clustering signals from read-out elements. The calorimeter information
consists of energy, position of entry point and time of arrival with
accuracies of $\sigma_{E}/E = 5.7\%/\sqrt{E(\mbox{GeV})}$, 
$\sigma_{z}=1.2\mbox{cm}/\sqrt{E(\mbox{GeV})}$,
$\sigma_{\phi}=\mbox{1.2 cm}$ and 
$\sigma_{t} = 54\mbox{ ps}/\sqrt{E(\mbox{GeV})} \bigoplus 50 \mbox{ ps}$. 
Calorimeter clusters not associated with a DC track 
indicate arrival of neutral particles and the computed time of flight
identifies photons with excellent precision. Time of flight also allows
good separation of electrons from muons, pions and kaons.
\\[5mm]
A superconducting coil surrounds the entire detector and produces a solenoidal 
field $B = 0.52$ T. 
The trigger \cite{TRIG} is based on the detection of isolated energy
deposits in the calorimeter and on hit multiplicity in the drift chamber. 
Only events triggered by the calorimeter have been used in the present
analysis. This choice ensures a far more reliable estimate of all necessary
efficiencies. 
The trigger system also includes a veto for cosmic-ray muons (cosmic ray
veto or CRV) based on energy deposits in the outermost layers of the
calorimeter and followed by a third-level software trigger able to identify
most of the \f\ events. 
A software filter (filfo or FLF), based on the topology and multiplicity
of calorimeter clusters and drift chamber hits, is applied 
to filter out machine background.
\\[5mm]
Both CRV and FLF are sources of event rejection. Their effect on the BR
measurement has been studied on control samples which do not undergo,
respectively, the CRV, and the FLF filter. 
%
\section{The measurement}
\begin{figure}[htb]
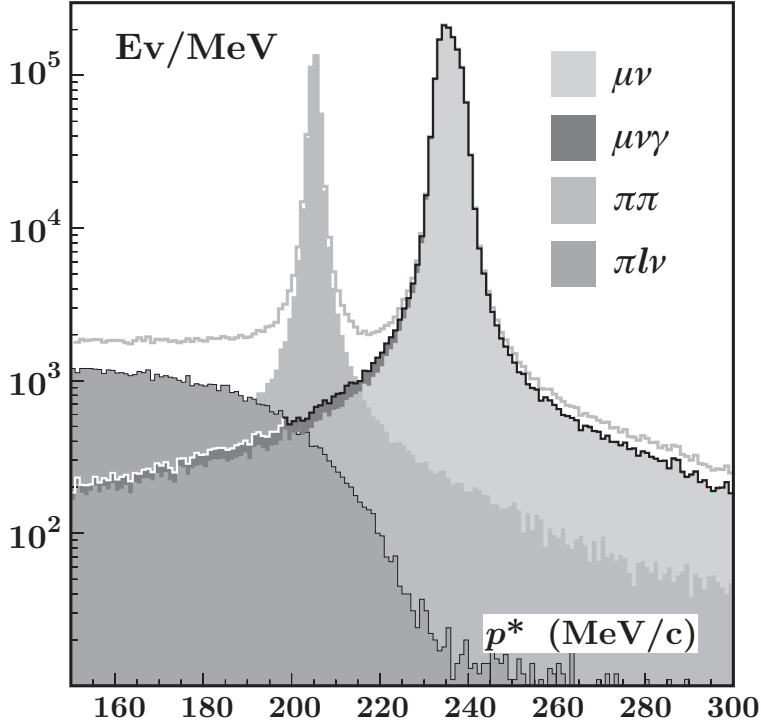

\figboxc newfig1;10;
\caption{Monte Carlo spectra of the charged decay particle momentum
  transformed to the kaon rest frame, assuming the pion mass. The two peaks
  corresponds to pions from \kppipi\, at 205 MeV/c and
  \kpmunu\ at 236 MeV. The black/white line represents the signal while 
  the grey line the signal plus background. Contributions from \kmuiig\
  are also shown.} 
\label{fig:mcspectrum}
\end{figure}

\begin{figure}[hptb]
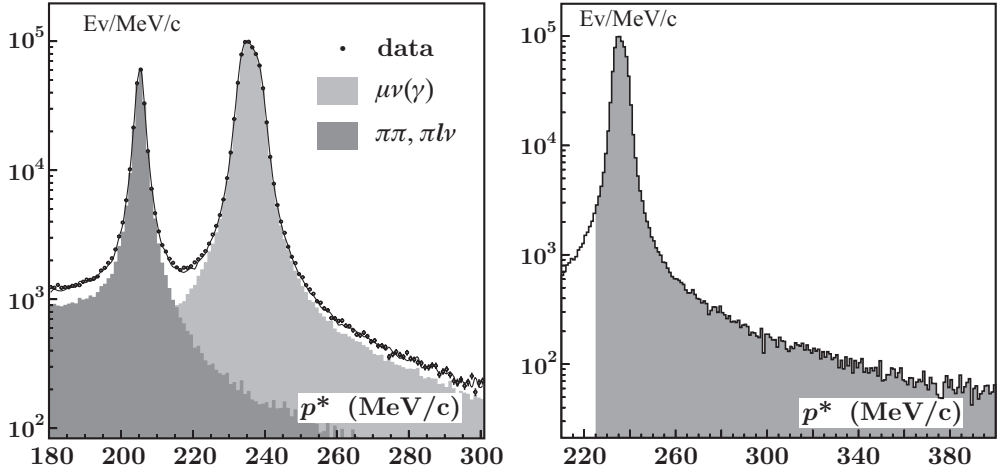

\cl{\figbox fig2a;6.6;\kern2mm\figbox fig2b;6.6;}
\caption{Left: Spectrum of the charged-secondary momentum in the kaon rest
  frame for data. The solid line indicates the fit done with a linear
  combination of signal and background distributions obtained from
  data. Right: The same spectrum after background subtraction. The range
  used for counting \kmuii\ events is indicated.} 
\label{fig:fitspectrum}
\end{figure}

The entire data sample of 175 pb\up{-1} is divided into two subsamples.\\ 
Some 60 pb\up{-1} of data have been used for the BR measurement. The
remaining 115 pb\up{-1} have been used to evaluate efficiencies and the
background. 
The branching ratio measurement \cite{kmu2note} is based on the use of
\ktagmuii\ decays for event tagging and to search for the \kmuii\
signal among $K^+$ decays. The tagging selection is based on the presence
of a two-tracks vertex in the DC which signals the $K^-$ decay \cite{kmu2note}. 
\\[5mm]
Nuclear interactions, NI, of kaons affects the branching ratio measurement
but not the tagging procedure. 
Since \sig\dn{\rm NI}(\kp)\ab\sig\dn{\rm NI}(\km)/10\up2, the choice above
minimizes the corrections to be calculated to account for the effect. The
corrections are in fact negligible.  
The large number of $K^+$ decays is sufficient for achieving a statistic
accuracy at the 0.1\% level, comparable with the systematic error.  
\\[5mm]
To avoid any bias due to differences in the trigger efficiency among the
$K^+$ decay modes on the signal ``hemisphere'', the
particles on the tagging side are required to deposit enough energy in the
calorimeter 
to trigger the data acquisition.
Nevertheless, the tagging criteria exhibit a residual, small dependence on
the decay mode of the $K^{+}$ on the signal hemisphere, introducing a tag
bias or TB, that has been studied using Monte Carlo simulation (MC) samples
\cite{offline} and checked on data. 
\\[5mm]
The search for positive kaon moving
outwards in the DC, with momentum $70< p_{K}<130$ Mev/c, is performed on
the sample of tagged events. The point of closest approach to the beam line
with coordinates $\{x,\,y,\,z\}$ 
is evaluated extrapolating the kaon track backwards to the beam line,  
taking into account the kaon energy loss. 
Kaon tracks with $|z|<20$ cm and $\sqrt{x^2+y^2}<$10 cm, and kaon decay
vertices in the fiducial volume, $40<\sqrt{x_V^2+y_V^2}<150$ cm, are selected. 
\\[5mm]
The number of \kmuii\ decays is obtained counting
the events with 225$\le p^{\ast}\le$400 MeV/c; 
$p^{\ast}$ is the charged decay particle momentum computed in the kaon
rest frame assuming the pion mass. 
The $p^{\ast}$ distribution is shown in fig. \ref{fig:mcspectrum}. 
\\[5mm]
%
%
The spectrum in fig.\ref{fig:mcspectrum}, obtained from MC simulation, shows a 2\%
contamination from various background sources, namely \kppipi , \kppil.
Since the maximum momentum of the pions from three-pion decays is 125
MeV/c, these channels do not contribute to the background. 
All the background sources in this analysis have one neutral pion 
in the final state. The neutral pions are identified by detecting the
photons from $\pi^{0}\to\gamma\gamma$ decay. The photons are identified as
isolated energy deposits in EMC not associated with tracks and satisfying
the constraints on  $\pi^0$ mass reconstruction and  
time correlation with the kaon decay vertex \cite{kmu2note}. This selection
allows us to obtain directly from data the $p^{\ast}$ distribution of the
background. \\[5mm]
The  $p^{\ast}$ distribution for the signal events has been
obtained from a data control sample described in the following efficiency evaluation.
This distribution has been used
together with the shape of the background sources to fit the overall
$p^{\ast}$ spectrum, figure \ref{fig:fitspectrum} left,  and to perform 
background subtraction.  
The result, after background subtraction, is shown in 
figure \ref{fig:fitspectrum} right. 
\\[5mm]
The branching ratio is obtained from:
\begin{equation}
 {\rm BR}(\kmuii)={N_{\kmuii}\over N_{Tag}}\x{1\over\:\epsilon\:C_{\rm
 CRV}\:C_{\rm FF}\:C_{\rm TB}} 
\label{eq:master}
\end{equation}
where $N_{\kmuii}$ are the events selected as signal (hereafter signal
count), $N_{Tag}$ is the number of tagged events and $\epsilon $ is the
efficiency.  
$C_{\rm CRV}$ and $C_{\rm FF}$ are the corrections for the effects 
due to the cosmic-ray veto and the filfo procedure, and  
$C_{\rm TB}$ accounts for the tag bias effect. 
\\[5mm]
%
%
The efficiency of the analysis cuts has been determined directly on data 
using a control sample of \kmu\ events selected exploiting their typical
signature in the EMC.
The control sample consists of events with $K^-\to\mu^-\nu(\gamma)$, 
providing the tag, and signal events \kmuii\  selected using only EMC
information.
This criterion is mostly independent from the selection procedure based
on DC information that has been used for obtaining the signal count.
The EMC selection requires only one cluster with E $>$ 80 MeV (High Energy
Cut), plus any number of clusters with 
energy below 20 MeV (Low
Energy Cut) which can be due to photons from \kmuiig\ . 
Further cuts on the energy ($E_{\rm cl}$) and on the distance in the tranverse
plane from the $z$-axis ($R_{xy}$)
of the cluster are applied to get rid of machine background and spurious
clusters.
Namely we require the cluster either to be on the barrel $R_{xy}>$197 cm or
to satisfy the relation $E_{\rm cl}/({\rm 1 GeV}) + R_{xy}/{\rm 5 (cm)} \ge 110$.
A correction of about $0.1 \%$, due to a tiny difference between
the efficiency evaluated on the control sample and 
the selection efficiency on the signal sample, 
has been estimated from MC.
The efficiency is $\epsilon= 0.3153 \pm 0.0002$. 
\\[5mm]
%
%
The corrections to  the event rejection described above, $C_{\rm CRV}$=1.0005 and 
$C_{\rm FF} - 1={\mathcal O}(10^{-5})$, have been directly measured on 
control samples which do not undergo, respectively, the cosmic-ray veto, 
and the filfo filter.
The correction for the tag bias, $C_{\rm TB}$=1.0164\plm0.0002, 
has been evaluated on Monte Carlo samples and the distribution of the
variables used for the tag selection have been checked on data.
%
%
The following sources of systematic uncertainties have been studied
varying the selection cuts: 
\begin{itemize}
\item [-]{the requirements on the tagging hemisphere;}
\item [-]{the trigger requirements;} 
\item [-]{the definition of the fiducial volume;}
\item [-]{the background evaluation procedure;}
\item [-]{the choice of the $[p^*_{min},p^*_{max}]$ range;}
\item [-]{the energy cuts for the efficiency sample;}
\item [-]{the effect of high-energy radiative photons ($E_\gamma>$20 MeV).}
\end{itemize}
Furthermore, the dependence of the measurement on the charged kaon 
lifetime, which affects 
the estimate of the geometrical acceptance,
has been studied varying the 
lifetime value used in the MC simulation.
The effects due to $K^+$ nuclear interactions have been evaluated from MC
simulation and measurements available in literature.
The stability of the measurement with respect to different data taking conditions
has been checked. 
The corresponding systematic errors are listed in table \ref{tab:errors}. \\
The statistical error due to the event count is $6\times 10^{-4}$ and
becomes $9\times 10^{-4}$ including the statistics of MC simulation and data used
for the efficiency evaluation.
\begin{table}[htbp]
\begin{center}
\begin{tabular}{||c|c||c|c||}
\hline
Source& Value\\
\hline
Low Energy Cut  & \pt5,-4,\\ \cline{1-2}
$E_\gamma>$20 MeV&\pt7,-4,\\ \cline{1-2}
High Energy Cut & \pt2,-4,\\ \cline{1-2}
Fiducial Volume & \pt5,-4,\\ \cline{1-2}
Background      &  \pt3,-4,\\ \cline{1-2}
$p^*$ range     &  \pt3,-4,\\ \cline{1-2}
Tag definition  & \pt1,-4,\\ \cline{1-2}
MC\ Lifetime    &$<\,$10\up{-6}\\ \cline{1-2}
Nuclear interactions&$<\,$\pt4,-4, \\ \cline{1-2}
Filfo           &$<\,$\pt3,-4,\\ \cline{1-2} 
Cosmic ray veto &$\mathcal{O}(10^{-6})$ \\ \cline{1-2}
Trigger &\pt9,-4,\\ \cline{1-2}
Total syst. & \pt15,-4, \\ \cline{1-2}
   \multicolumn{2}{c}{} \\
\end{tabular}
\end{center}
\caption{Summary table of systematic uncertainties.} 
\label{tab:errors}
\end{table}
%
%
\section{Conclusions}
\subsection*{BR Measurement}
On a sample of tagged events $N_{Tags}=4,237,329$, 
we found a number of signal events,
with $ 225 \le p^{\ast} \le 400$ MeV/c,
 of $N_{\kmuii}=865,283$. \\    
\noindent
Using eq. \ref{eq:master}, 
the absolute branching ratio is:
\begin{equation}
   BR (\kmuii) =
   0.6366 \pm 0.0009_{stat.} \pm 0.0015_{syst.}
   \label{eq:brgold}
\end{equation}
The KLOE measurement of the BR(\kmuii), fully inclusive of
final-state radiation, has a 0.27\% accuracy and it is based on an
unprecedented statistics and carefully controlled systematics. 
%
\subsection*{$V_{us}$ extraction}
The recent publication \cite{MILC} of the results of lattice QCD calculations 
has renewed the interest in improving the accuracy of the BR(\kmuii),
which represents an experimental alternative to the semileptonic kaon decays 
in measuring $|\vus|$ as pointed out by Marciano in ref. \citen{Marciano}.
\noindent
The extraction of this CKM matrix element
is based on the ratio of the decay rates 
for the inclusive decays \kmuii\ 
and $\pi^{+}\rightarrow\mu^{+}\nu(\gamma)$: 
\begin{equation}
{\Gamma(K\to\mu\nu(\gamma))\over\Gamma(\pi\to\mu\nu(\gamma))}=
{\:m_K\left(1-{m_\mu^2\over m_K^2}\right)^2
\over \:m_\pi\left(1-{m_\mu^2\over m_\pi^2}\right)^2}\:
 \frac{\:|\vus|^2\:} {\:|\vud|^2\:}
\:\frac{\:f_K^2\:}{\:f_\pi^2\:}
\:\frac{\:1 + \frac{\alpha}{\pi} C_K\:}
             {\:1 + \frac{\alpha}{\pi} C_\pi\:}
\label{eq:ratio1}
\end{equation}
where $f_K$ and $f_\pi$ are, respectively, the kaon and the pion decay
constants; $C_\pi$ and $C_K$ parametrize the radiative-inclusive electroweak
corrections, 
taking into account bremsstrahlung emission of real photons and of 
virtual-photon loop contributions as well.
\noindent
Using the branching ratios of \kmuii\ and \pmuii\ decays, 
the $|\vud|$ value from super-allowed nuclear beta decays,
$C_\pi$ and $C_K$ from ref. \citen{Marciano} and references therein,
 and 
the new lattice calculation 
of $f_K / f_\pi$ \cite{MILC}, it is possible to 
extract $|\vus|$ with an uncertainty at the percent level, whose error is
mainly dominated by the accuracy of   
lattice calculations.
From the BR measurement, using the determination of
 $f_K / f_{\pi}$ = 1.210(4)(13), we have obtained the ratio:
\begin{equation} 
\left | \frac{V_{us}}{V_{ud}} \right |^2 = 0.05211\pm0.00016\pm0.00019\pm0.00117
   \label{eq:ckmratio}
\end{equation}
\noindent
where the errors correspond, respectively, to the experimental, the
structure-dependent radiative corrections, and the lattice uncertainties.
\noindent
Taking $V_{ud}$ from super-allowed nuclear $\beta$ decays, 
$V_{ud}=0.9740\pm0.0005$ \cite{Czarnecki:2004cw}, one determines $|V_{us}|$:
\begin{equation}
  |V_{us}|_{\kmuii} =  0.2223\pm0.0026 .
   \label{eq:valvus}
\end{equation}
The accuracy is dominated by the knowledge of  $f_K / f_{\pi}$ from lattice
calculation.
Alternatively, the unitarity relationship  $|V_{ud}|^2 = 1 - |V_{us}|^2$ can
be assumed in eq. \ref{eq:ckmratio} giving: 
\begin{equation}
  |V_{us}|_{\rm unitarity} =  0.2225\pm0.0025
   \label{eq:univus}
\end{equation}   
independent of the $V_{ud}$ measurement.
\noindent
The result quoted in eq. \ref{eq:valvus} is in agreement with the
unitarity of the CKM matrix (eq. \ref{eq:univus})  and with the 
determinations of $V_{us}$ from semileptonic decays, whose precision is
dominated by the $f_+(0)$ calculations \cite{f0ref}.
\section*{Acknowledgements}
We thank the DA$\Phi$NE team for their efforts in maintaining low background 
running conditions and their collaboration during all data-taking. 
We want to thank our technical staff: G. F. Fortugno for his dedicated work
to ensure an efficient operation of the KLOE Computing Center; 
M. Anelli for his continous support to the gas system and the safety of the
detector; A. Balla, M. Gatta, G. Corradi and G. Papalino for the maintenance of
the electronics; M. Santoni, G. Paoluzzi and R. Rosellini for the general 
support to the detector; C. Piscitelli for his help during major maintenance 
periods.
\noindent
This work was supported in part by DOE grant DE-FG-02-97ER41027; 
by EURODAPHNE, contract FMRX-CT98-0169; by the German Federal Ministry of
Education and Research (BMBF) contract 06-KA-957; by Graduiertenkolleg
`H.E. Phys. and Part. Astrophys.' of Deutsche Forschungsgemeinschaft,
Contract No. GK 742; by INTAS, contracts 96-624, 99-37; and by TARI,
contract HPRI-CT-1999-00088. 
%
%

%
\end{document}